\documentclass[%
 aip,
 jmp,%
 amsmath,amssymb,
preprint,%
]{revtex4-1}

\usepackage{graphicx}
\usepackage{dcolumn}
\usepackage{bm}
\usepackage{float}
\usepackage{color}
\usepackage[draft]{changes} 
\usepackage[normalem]{ulem}
\definechangesauthor[name={yi}, color=blue]{Yi}
\definechangesauthor[name={Maria}, color=red]{Maria}
\newcommand{\argonneCNM}{Center for Nanoscale Materials, Argonne National Laboratory, Argonne, IL 60439, USA}
\newcommand{\argonneMSD}{Materials Science Division, Argonne National Laboratory, Argonne, IL 60439, United States}
\newcommand{\northwesternCHEM}{Department of Chemistry, Northwestern University, Evanston, IL 60208, United States}

\begin{document}
\title{Lattice thermal transport in group II-alloyed PbTe}

\author{Yi Xia} 
\email{yxia@anl.gov}
\affiliation{Center for Nanoscale Materials, Argonne National Laboratory, Argonne, Illinois 60439, USA}
\author{James M. Hodges}
\affiliation{Department of Chemistry, Northwestern University, Evanston, IL 60208, United States}
\author{Mercouri G. Kanatzidis}%
\affiliation{Department of Chemistry, Northwestern University, Evanston, IL 60208, United States}
\affiliation{Materials Science Division, Argonne National Laboratory, Argonne, IL 60439, United States}
\author{Maria K Y Chan}
\email{mchan@anl.gov}
\affiliation{Center for Nanoscale Materials, Argonne National Laboratory, Argonne, Illinois 60439, USA}

\date{\today}

\begin{abstract}
PbTe, one of the most promising thermoelectric materials, has recently demonstrated thermoelectric figure of merit ($ZT$) of above 2.0 when alloyed with group II elements.  The improvements are due mainly to significant reduction of lattice thermal conductivity ($\kappa_{l}$), which was in turn attributed to nanoparticle precipitates. However, a fundamental understanding of various phonon scattering mechanisms within the bulk alloy is still lacking. In this work, we apply the newly-developed density-functional-theory (DFT)-based compressive sensing lattice dynamics (CSLD) approach to model lattice heat transport in PbTe, MTe, and Pb$_{0.94}$M$_{0.06}$Te (M=Mg, Ca, Sr and Ba), compare our results with experimental measurements, with focus on strain effect and mass disorder scattering. We find that (1) CaTe, SrTe and BaTe in the rock-salt structure exhibit much higher $\kappa_{l}$ than PbTe, while MgTe in the same structure shows anomalously low $\kappa_{l}$; (2) lattice heat transport of PbTe is extremely sensitive to static strain induced by alloying atoms in solid solution form; (3) mass disorder scattering plays a major role in reducing $\kappa_{l}$ for Mg/Ca/Sr-alloyed PbTe through strongly suppressing the lifetimes of intermediate- and high-frequency phonons, while for Ba-alloyed PbTe, precipitated nanoparticles are also important.
\end{abstract}

\maketitle

Thermoelectric (TE) devices, which are capable of converting waste heat into electricity, are ideal alternative renewable energy technologies to overcome limited fossil fuel resources and environmental challenges.\cite{Sootsman2009review} Thermoelectric energy conversion efficiency is characterized by the dimensionless figure of merit $ZT=S^2\sigma T/(\kappa_{e}+\kappa_{l})$, where $S$, $\sigma$ and $T$ are the Seebeck coefficient, the electrical conductivity, and temperature, and $\kappa_{e}$ and $\kappa_{l}$ are the electronic and lattice thermal conductivities, respectively. $ZT$ is generally optimized by maximizing the thermoelectric power factor $S^2\sigma$ and minimizing $\kappa_{l}$.\cite{Gangjian2016,Lidong2014review} 

PbTe-based TE materials are among the highest performing, partly because of strong inherent phonon anharmonicity, leading to low $\kappa_{l}$.\cite{Zhang201592}  To further enhance $ZT$, simultaneously reducing $\kappa_{l}$ and improving the power factor can be achieved by introducing anionic and cationic dopants. Recent experiments show that $p$-type PbTe alloyed with group II elements (Mg, Ca, Sr and Ba) achieved $ZT$ well above 1.5.\cite{Biswas2011,Biswas2011EES,Biswas:2012aa,Zhao2013,Tan2016,Fu2016141,Ohta2012} It is found in experiments and theoretical calculations that, with group II dopants, electronic properties are improved owing to convergence of multiple valence bands and band gap widening.\cite{Zhao2013,Tan2016} With respect to lattice heat transport, significant reduction of $\kappa_{l}$ is observed, which was attributed to the all-scale hierarchical architectures-induced phonon scattering due to the presence of solid-solution point defects, nanoscale precipitates, and grain boundaries.\cite{Biswas:2012aa,Zhang201592} However, fundamental understanding of the roles of various phonon scattering mechanisms in reducing $\kappa_{l}$ is still lacking. Previous first-principles investigations mainly focused on pristine PbTe and its anionic alloys (PbSe$_{x}$Te$_{1-x}$),\cite{zhiting,Lee:2014aa} showing the large intrinsic anharmonicity and the importance of optical phonons which provide strong scattering channels for acoustic phonons. Recently, temperature-induced phonon renormalization was further investigated using quasiharmonic approximation and temperature-dependent effective potential techniques.\cite{Skelton2014,Romero2015} However, lattice thermal transport for group II alloyed-PbTe is not well understood, with existing theoretical modeling limited to phenomenological Debye-Callaway model.\cite{Jiaqing2010,Lo2012,Tan2016} Generally, various scattering mechanisms are present in the alloyed phases including intrinsic anharmonic scattering and extrinsic scatterings from precipitates, strains and point defects.\cite{Lo2012} It is fundamentally important to understand the role of various phonon scattering mechanisms in order to explore the possibility to further reduce $\kappa_{l}$. Motivated by this point, we use first-principles methods and recently-development compressive sensing lattice dynamics (CSLD)\cite{csld} approach to investigate lattice heat transport of PbTe, MTe, and Pb$_{0.94}$M$_{0.06}$Te (M=Mg, Ca, Sr and Ba), particularly focusing on the effects of strain and alloying on $\kappa_{l}$.

The lattice thermal conductivity tensor ($\kappa_{l}^{\alpha\beta}$) was calculated by summing over contributions from phonon modes in the first Brillouin zone under the relaxation time approximation,\cite{ziman}
	\begin{equation}\label{eq:kappa}
	\kappa_{l}^{\alpha\beta}=\frac{1}{k_{B}T^2\Omega N} \sum_{\lambda} f_{0} (f_{0}+1) (\hbar \omega_{\lambda})^2 v_{\lambda}^{\alpha} v_{\lambda}^{\beta} 	\tau_{\lambda},
	\end{equation}
where $N$, $\Omega$ and $f_{0}$ are the number of phonon wave vectors, the volume of primitive cell, and the Bose-Einstein distribution function, respectively. For each mode $\lambda$, $\omega_{\lambda}$, $v_{\lambda}^{\alpha}$ and $\tau_{\lambda}$ represent the phonon frequency, the velocity along $\alpha$ direction, and the mode lifetime, respectively. The lifetime is calculated using Fermi's golden rule by treating $3^{\text{rd}}$-order interatomic force constants as perturbations to harmonic phonons,\cite{ziman} and the linearized Boltzmann transport equation (BTE) is solved in an iterative manner to take into account non-equilibrium phonon distributions.\cite{omini1,omini2,broido,Ward2009,wuli} Effects of alloying on the phonon spectra are treated by taking into account extrinsic phonon scattering due to mass disorder,\cite{Abeles1963} which is treated as an external elastic scattering in BTE. The associated phonon mode scattering rates are given by
	\begin{eqnarray}
	\Gamma_{\lambda\lambda^{'}} &=& \frac{\pi\omega^2}{2} \sum_{a} g(a) \left |  \mathbf{e}_{a}^{\lambda*} \cdot \mathbf{e}_{a}^{\lambda^{'}}  \right |^2 \delta(\omega_\lambda-\omega_{\lambda'}),\\
	g(a) &=& \sum_{s} x_{s}(a) [1-M_{s}(a)/\bar{M}(a)]^2,
	\end{eqnarray}
where $x_{s}(a)$ and $M_{s}(a)$ are, respectively, the concentration and the atomic mass of the $s$-th species on the atomic site indexed by $a$. 
The ShengBTE package\cite{shengbte} was used to perform the iterative calculation of phonon lifetime and $\kappa_{l}$ with 24$\times$24$\times$24 $\mathbf{q}$-point meshes.	

Both $2^{\text{nd}}$- and $3^{\text{rd}}$-order interatomic force constants (IFCs) were extracted using the recently-developed compressive sensing lattice dynamics (CSLD).\cite{csld} CSLD belongs to the class of direct supercell methods where forces  calculated using density functional theory (DFT) are used to parametrize the force-displacement relation via Taylor expansion.\cite{csld} By taking advantage of the sparsity of the space of IFCs and  crystal symmetry, CSLD is an efficient method to extract high-order IFCs. Its accuracy has been verified in both lattice dynamics simulations and cluster expansion models.\cite{csce1, csce2, Jiangang2016,Jiangang2017,vo2yi} (see Supplemental Materials for detailed discussion of IFCs fitting and computational parameters)

\begin{figure*}[htp]
	\includegraphics[width = 0.95\linewidth]{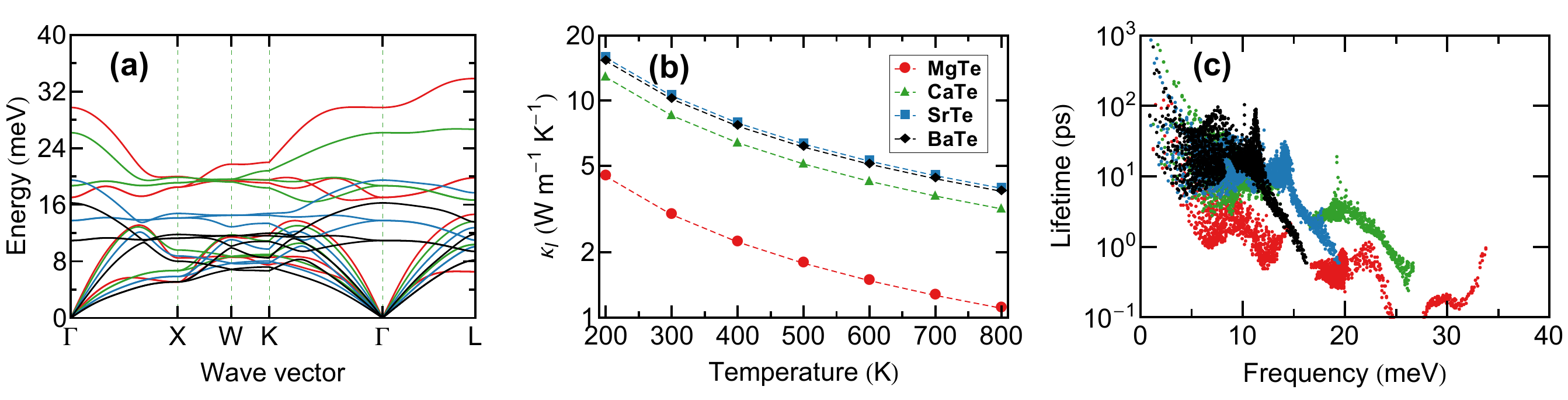}
	\caption{ (a) Phonon dispersion, (b) temperature-dependent lattice thermal conductivity and (c) phonon mode lifetime at 300 K for rock-salt MTe (M=Mg, Ca, Sr and Ba), which are colored in red, green, blue and black respectively. }
	\label{fig:MTe}
\end{figure*}

\begin{figure*}[htp]
	\includegraphics[width = 0.95\linewidth]{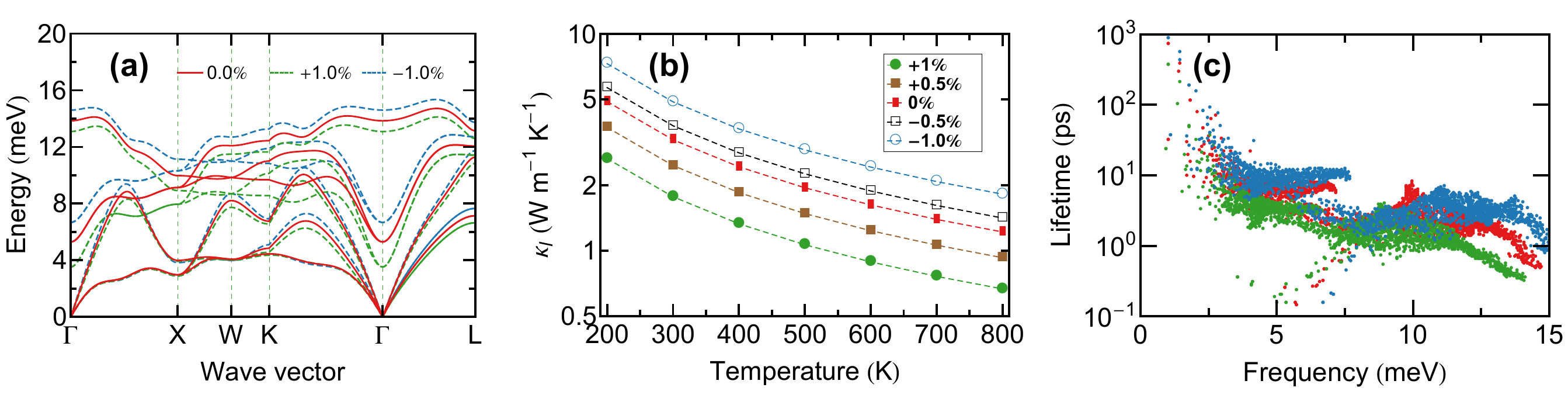}
	\caption{ (a) Phonon dispersions for PbTe with lattice parameter fully relaxed (red), contracted (dashed blue) and expanded 
	(dashed green) by 1\% respectively. (b) Lattice thermal conductivity of PbTe as a function of temperature with lattice parameters
	 varied from -1\% to 1\%. (c) Phonon mode lifetime with 0 (red), 
	 -1\% (blue) and 1\% (green) lattice expansion at 300 K. }
	\label{fig:Strain}
\end{figure*}

Owing to the general low solubility of MTe in PbTe matrix, MTe are experimentally found to precipitate out of the matrix, forming both coherent and incoherent nanoscale precipitates.\cite{Biswas2011,Biswas:2012aa,Zhao2013,Tan2016,Fu2016141} Since $\kappa_{l}$ reduction is suggested to originate from these precipitates, it is useful to compare $\kappa_{l}$ of pristine PbTe and MTe. To shed light on this point, phonon dispersion and $\kappa_{l}$ were computed and analyzed. The phonon dispersions shown in Figure\ \ref{fig:MTe}(a) share similar features and exhibit enhanced phonon softening with increased atomic mass from Mg to Ba. Noticeable difference is observed in MgTe, which shows extra softening of acoustic modes near the $X$ and $L$ point in the first Brillouin zone. Compared to the phonon dispersion of PbTe in Figure\ \ref{fig:Strain}(a), one important feature of MTe is the absence of low-lying transverse optical (TO) phonons,  which are found to be essential to scatter acoustic phonons,\cite{zhiting} thus making their $\kappa_{l}$ higher than PbTe. The computed values of $\kappa_{l}$ for MgTe, CaTe, SrTe, BaTe (Figure\ \ref{fig:MTe}(b)) and PbTe (Figure\ \ref{fig:Strain}(b)) are 3.0, 8.5, 10.5, 10.2 and 3.3 W/m$\cdot$K at room temperature using fully relaxed lattice parameter at 0 K. The value for PbTe is higher than previously reported computed values (1.9 and 2.1 W/m$\cdot$K).\cite{zhiting,Skelton2014} The discrepancy may be attributed to two factors: (1) $\kappa_{l}$ of PbTe is extremely sensitive to lattice parameter (we will further illustrate this point by considering strain effect in the next section), and therefore can be heavily influenced by the adopted xc functional; (2) $\kappa_{l}$ is potentially underestimated under the single mode relaxation time approximation (SMRTA) adopted by the previous studies, which, according to our calculations, is found to give a value of $\kappa_{l}$ lower by 0.6 W/m$\cdot$K compared to the more accurate treatment of iterative solution of the BTE. Consistent with the difference found in phonon dispersion between MTe and PbTe, $\kappa_{l}$ is higher for CaTe, SrTe and BaTe than for PbTe at 300 K. However, MgTe, despite having the lightest element, exhibits the lowest $\kappa_{l}$ among MTe, with a value comparable to that of PbTe. Since the two main factors that determine $\kappa_{l}$ are lifetimes and group velocities, we plotted the computed mode-dependent lifetimes in Figure\ \ref{fig:MTe}(c). It can be seen that MgTe has much shorter lifetimes compared with the others in the entire range of vibrational frequency, which explains the significantly reduced $\kappa_{l}$. To further shed light on the anomalously low $\kappa_{l}$ of rock-salt MgTe, we also computed the $\kappa_{l}$ of the more stable zinc-blende phase of MgTe, and found a value of 14.1 W/m$\cdot$K at 300 K. Detailed analysis shows that although the magnitude of 3$^{\text{rd}}$-order IFCs are reduced with structure change from zinc blende to rock salt, the allowed three-phonon processes is significantly enhanced (nearly twice in the absorption process) with the more suppressed optical modes in the rock-salt structure, giving rise to anomalously low $\kappa_{l}$. The fact that all MTe compounds have larger or similar $\kappa_{l}$ compared with PbTe indicates that extrinsic phonon scattering from interfaces, point defects, grain boundaries, and precipitates should be examined to explain the reduced $\kappa_{l}$ of group II-alloyed PbTe.

Aside from nanoscale precipitates, the size and strain fluctuations caused by point defects in the solid solution could also play an important role in determining the overall $\kappa_{l}$ in alloyed PbTe. The fully relaxed lattice parameters of MgTe, CaTe, SrTe and BaTe are 5.971, 6.400, 6.723, and 7.093 \r{A} respectively. When PbTe (fully relaxed lattice parameter of 6.441 \r{A}) form solid solutions with 6\% (12\%) MTe, its lattice parameters thus could vary roughly from -0.5\% (-1.0\%) to 0.5\% (1.0\%). It was shown in quasiharmonic approximation calculation\cite{Skelton2014} that $\kappa_{l}$ is very sensitive to PbTe lattice parameter. However, the effects of static strain, specifically caused by alloying with group-II elements, on the $\kappa_{l}$ of PbTe was not taken into account in previous studies,\cite{Jiaqing2010,Lo2012,Tan2016} motivating us to investigate its effect on phonon dispersion and $\kappa_{l}$. Figure\ \ref{fig:Strain}(a) shows the phonon dispersion of PbTe calculated with -1\%, 0\%, and 1\% lattice expansion respectively. One striking feature is the presence of strong softening of optical phonon modes with lattice expansion. Particularly, TO modes near the $\Gamma$ point fall deep into the acoustic region when the lattice is slightly expanded. The increased overlap between TO and acoustic modes increases the scattering phase space,\cite{Romero2015} giving rise to further reduction in $\kappa_{l}$. The corresponding $\kappa_{l}$ plotted in Figure\ \ref{fig:Strain}(b) shows that $\kappa_{l}$ can be increased from 3.3 to 3.8 (4.8) W/m$\cdot$K with 0.5 (1.0)\% lattice contraction, and decreased to 2.5 (1.8) W/m$\cdot$K with 0.5 (1.0)\% lattice expansion. Since the relaxed lattice parameter of 6.441 \r{A} is smaller than the experimental value of 6.462\r{A} at 300 K,\cite{Kastbjerg2013} we computed $\kappa_{l}$ using the experimental lattice parameter and found a value of 2.7 W/m$\cdot$K (2.1 W/m$\cdot$K under SMRTA) at 300 K. Our results compare very well with Ref.\ \onlinecite{Skelton2014} which used the same xc functional, and our higher value is from solving BTE in an iterative (more accurate) manner. Compared to experiments, our result is close to measurement performed on single-crystal PbTe of 2.4 W/m$\cdot$K.\cite{Morelli2008} The slight overestimation can be attributed to: (1) overestimation of the energy of low-lying TO optical phonon modes (see Fig.~S2 in Supplementary Materials), and (2) neglect of higher-than-third-order phonon-phonon interactions. We note that experimental measurements on polycrystalline samples yield a yet lower $\kappa_{l}$ of 2.0 W/m$\cdot$K,\cite{El-Sharkawy1983} due to additional phonon scattering from defects and grain boundaries. Despite of the overestimation presented in pristine PbTe , our result is close to $\kappa_{l}$ of 2.8 W/m$\cdot$K in Na-doped PbTe, which allows us to further investigate the additional alloying effects from group II elements. Our results also reveal that strain effect alone can significantly influence the intrinsic anharmonicity of PbTe matrix and different effects should be expected for different group II elements. By examining the phonon lifetime in Figure\ \ref{fig:Strain}(c), we find that lattice parameter increase tends to enhance the scattering rate, revealing that lifetime reduction plays a major role in reducing $\kappa_{l}$ with lattice expansion. Note that here we only consider the static strain effect and ignore phonon scattering due to spatial strain fluctuations.

\begin{figure*}[htp]
	\includegraphics[width = 0.98\linewidth]{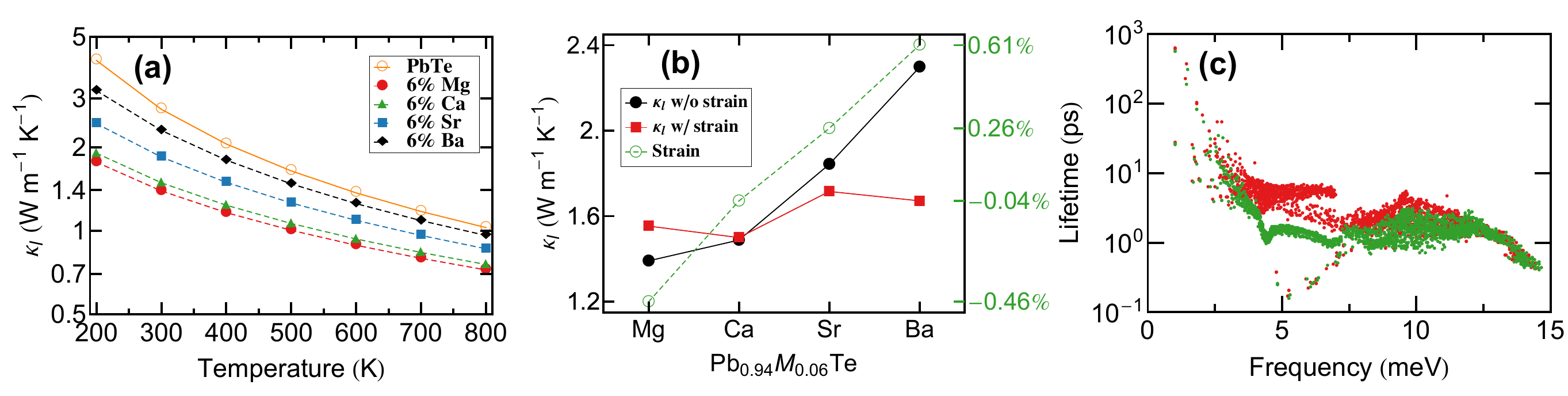}
	\caption{ (a) Temperature-dependent lattice thermal conductivity of PbTe, and alloyed with 6\% MgTe, CaTe, SrTe and BaTe respectively.
	(b) Lattice thermal conductivity of Pb$_{0.94}$M$_{0.06}$Te (M=Mg, Ca, Sr and Ba) with and without strain effect (black and red respectively) at 300 K. The corresponding lattice expansion/contraction (green)  is shown on the right vertical axis.
	(c) Phonon mode lifetime of pristine PbTe (red) and Pb$_{0.94}$Mg$_{0.06}$Te (green) at 300 K. }
	\label{fig:Mass}
\end{figure*}

Considering the strain effect only, the $\kappa_{l}$ of PbTe should increase with MgTe alloying and decrease with SrTe alloying. However, experimentally $\kappa_{l}$ decreases with both group II alloys, indicating that additional scattering mechanisms may dominate over the strain effect. Another important factor introduced by alloying elements is mass fluctuations in crystals. By neglecting the change and disorder of IFCs induced by MTe and assuming random distribution of M atoms over Pb sites, we examine the phonon scattering due to mass disorder, i.e., alloy scattering. To better illustrate the effect of mass disorder scattering, here we use experimental lattice parameter of PbTe which shows better agreement between theoretical and experimental $\kappa_{l}$. Computed $\kappa_{l}$ of PbTe alloyed with 6\% MTe are shown in Figure\ \ref{fig:Mass}(a). We find that the largest $\kappa_{l}$ reduction can be achieved through alloying with MgTe which has the largest mass contrast, while BaTe with the most similar mass as PbTe reduces $\kappa_{l}$ much less effectively. This is due to the fact that mass disorder scattering rates are proportional to mass contrast and increase with increased relative alloying concentration. Further considering strain effect on top of mass disorder scattering, as shown in Figure\ \ref{fig:Mass}(b) and Table~S1 and~S2 in Supplementary Materials, we find that $\kappa_{l}$ changes from 1.39, 1.49, 1.84 and 2.30 W/m$\cdot$K to 1.55, 1.50, 1.72, and 1.67 W/m$\cdot$K for Mg, Ca, Sr and Ba respectively, where the decrease/increase of $\kappa_{l}$ is in line with the corresponding lattice expansion/contraction. The strain effect on Pb$_{0.94}$Ba$_{0.06}$Te that leads to significant reduction of $\kappa_{l}$  indicates its importance when there is large lattice mismatch. Compared to experiments, our results of Pb$_{0.94}$Mg$_{0.06}$Te agree well with recent experimental report that the room temperature $\kappa_{l}$ is significantly reduced from 2.83 to 1.74 W/m$\cdot$K for Pb$_{0.94}$Mg$_{0.06}$Te.\cite{Fu2016141,Zhao2013} For Pb$_{0.94}$Ca$_{0.06}$Te, experimental value of 1.33 W/m$\cdot$K also agrees with our theoretical value of 1.50 W/m$\cdot$K at 300 K.\cite{Biswas2011EES} For Pb$_{0.94}$Sr$_{0.06}$Te, the computed $\kappa_{l}$ of 1.72 W/m$\cdot$K agrees reasonably with experimentally reported value of about 2.0 W/m$\cdot$K,\cite{Tan2016} where the authors utilized non-equilibrium processing technique to extend the solubility from less than 1 mol\% to about 5 mol\%. The slight underestimation in theory could be due to the requirement of full solubility in modeling mass disorder scattering. Our results suggest that mass disorder scattering plays the most important role in reducing $\kappa_{l}$ for cationic dopants with large mass contrast (such as Mg, Ca and Sr) near or slightly exceeding the solubility limit, and relatively weakens the extra phonon scattering caused by static strain compared to pristine PbTe. Our results also suggest that extrinsic scattering, for example induced by precipitated nanostructures and nanostructuring which limits maximum mean free path, is not required to explain the experimental results when alloying group II atoms are mostly in solid solution form. For cationic dopant with reduced mass contrast such as Ba, however, $\kappa_{l}$ reduction cannot be primarily attributed to strain effect or mass disorder scattering. In contrast to Mg, Ca and Sr, nanostructures are expected to be responsible for $\kappa_{l}$ reduction with Ba. This can be inferred from (1) the large difference between the computed $\kappa_{l}$ of 1.67 W/m$\cdot$K for Pb$_{0.94}$Ba$_{0.06}$Te and experimentally observed $\kappa_{l}$ of about 1.20 W/m$\cdot$K in Pb$_{0.97}$Ba$_{0.03}$Te\cite{Biswas2011EES} and (2) Ba has extremely low solubility (less than 0.5\%) in PbTe.\cite{Lo2012} Comparison between phonon lifetime in Figure\ \ref{fig:Mass}(c) confirms the strong scattering of optical modes through introducing mass disorder, while acoustic modes are less affected, suggesting that $\kappa_{l}$ can be further reduced by creating mesoscale grain boundaries without significantly affecting electronic transport. Note that here we only compare to experimental results at 300 K since we only considered room-temperature lattice parameter. We expect that our conclusion remains qualitatively sound at high temperatures, where lattice expansion and phonon renormalization (hardening of TO modes) could further affect $\kappa_{l}$, though it may quantitatively varies.

We have applied a first-principles-based compressive sensing lattice dynamics approach for modeling lattice heat transport in PbTe and its solid solutions formed with group II elements. With the computed harmonic and anharmonic interatomic force constants, we modeled the lattice heat transport by solving the linearized Boltzmann transport equations with phonon lifetimes obtained from perturbation theory. Lattice thermal conductivities were computed and analyzed for PbTe and its alloys within the framework of virtual crystal approximation, with a particular emphasis on the effects of strain and mass disorder. The results show that the current approach is able to explain the experimentally observed reduction of lattice thermal conductivity, thus allowing further improvement of $ZT$ through phonon engineering.

\textbf{Acknowledgements} This work was supported by the Midwest Integrated Center for Computational Materials (MICCoM) as part of the Computational Materials Sciences Program funded by the U.S. Department of Energy, Office of Science, Basic Energy Sciences, Materials Sciences and Engineering Division (5J-30161-0010A). Use of the Center for Nanoscale Materials, an Office of Science user facility, was supported by the U. S. Department of Energy, Office of Science, Office of Basic Energy Sciences, under Contract No. DE-AC02-06CH11357. This research used resources of the National Energy Research Scientific Computing Center, a DOE Office of Science User Facility supported by the Office of Science of the U.S. Department of Energy under Contract No. DE-AC02-05CH11231. We gratefully acknowledge the computing resources provided on Blues, a high-performance computing cluster operated by the Laboratory Computing Resource Center at Argonne National Laboratory.

\bibliography{PbTe}

\widetext
\clearpage
\begin{center}
	\textbf{\large Supplementary Materials: Lattice thermal transport in group II-alloyed PbTe}
\end{center}

\begin{center}

Yi Xia$^{1}$, James M. Hodges$^{2}$, Mercouri G. Kanatzidis$^{2,3}$ and  Maria K. Y. Chan$^{1}$

\vspace{0.3cm}

\text{$^1$\argonneCNM}
\text{$^2$\northwesternCHEM}
\text{$^3$\argonneMSD}

\end{center}

\setcounter{equation}{0}
\setcounter{figure}{0}
\setcounter{table}{0}
\setcounter{page}{1}
\makeatletter
\renewcommand{\thepage}{S\arabic{page}}
\renewcommand{\theequation}{S\arabic{equation}}
\renewcommand{\thefigure}{S\arabic{figure}}
\renewcommand{\thetable}{S\arabic{table}}
\renewcommand{\bibnumfmt}[1]{[S#1]}
\renewcommand{\citenumfont}[1]{S#1}



\section{Computational details} 

The Vienna $Ab\ initio$ Simulation Package (VASP) \cite{Vasp1, Vasp2, Vasp3, Vasp4} was used to perform DFT calculations. The projector-augmented wave (PAW)\cite{paw} method was used in conjunction with the Perdew-Burke-Ernzerhof revised for solids (PBEsol)\cite{pbe,Perdew2008} generalized gradient approximation (GGA)\cite{gga} for the exchange-correlation (xc) functional.\cite{dft} For structural relaxation, Monkhorst-Pack $\mathbf{k}$-point meshes of 16$\times$16$\times$16 and a plane wave basis with a kinetic energy cutoff of 520 eV were used. The force and energy convergence threshold were set to be $10^{-3}$ eV/\r{A} and $10^{-8}$ eV respectively. Supercells (4$\times$4$\times$4) with randomly perturbed atomic positions from 0.01 \r{A} to 0.05 \r{A} were constructed to fit interatomic force constants (IFCs) using CSLD. The supercell structures were sampled with 2$\times$2$\times$2 $\mathbf{k}$-point meshes and computed using the same convergence settings. Non-analytic correction of phonon dispersion near Gamma point was performed using the mixed-space approach,\cite{Wang2010} with Born effective charges and macroscopic static dielectric constant computed by density function perturbation theory (DFPT).\cite{Baroni1986,Gajdo2006}

\section{Results and discussions}

\subsection{Interatomic force constant fitting using compressive sensing lattice dynamics}
\begin{figure}[htp]
	\includegraphics[width = 0.85\linewidth]{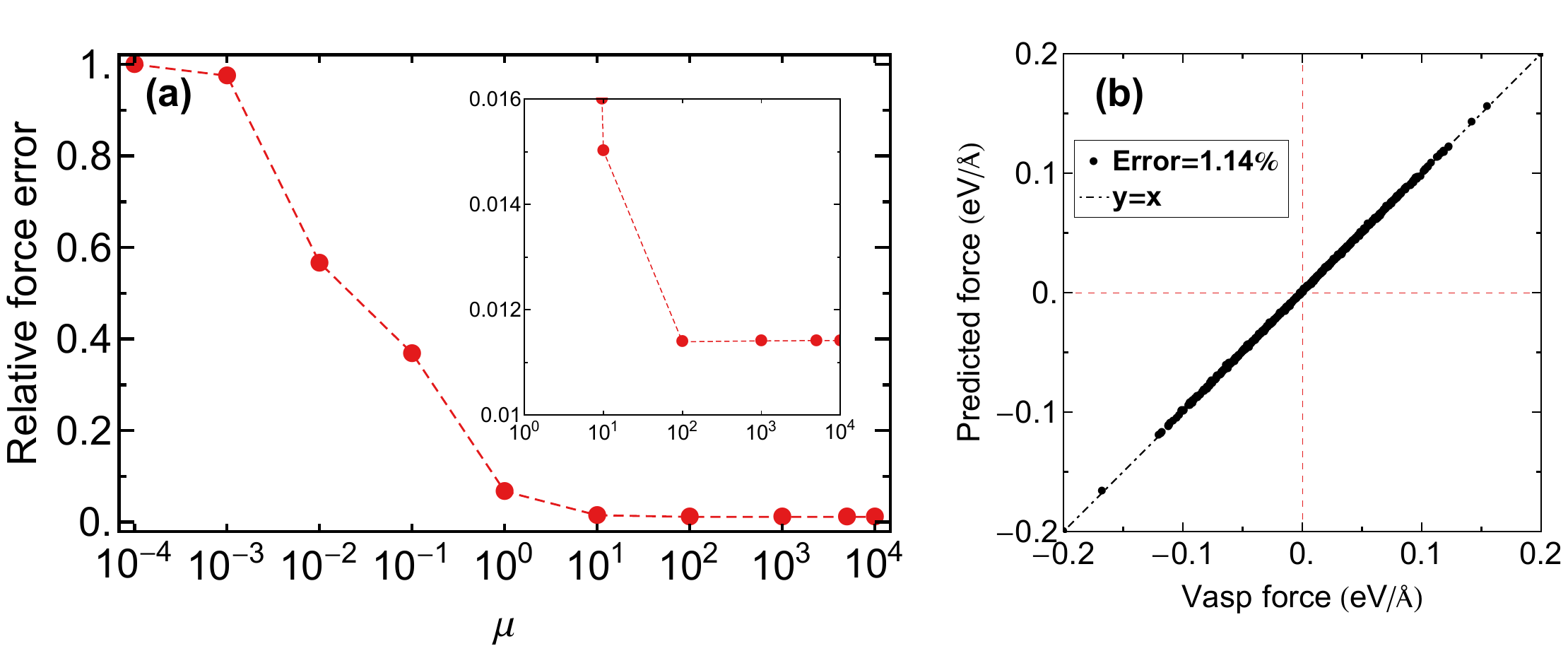}
	\caption{
	(a) Relative force error for a leave-out subset as a function of parameter $\mu$. The inset shows more details with $\mu$ varying from 10 to 10$^{4}$. (b) Predicted forces from CSLD compared with forces directly computed by VASP using optimal $\mu$, which leads to relative force error of 1.14\%.
	}
	\label{fig:fitcsld}
\end{figure}

As detailed in Ref.~\onlinecite{csld}, the IFCs ($\mathbf{\Phi}$) is obtained from a convex optimization problem that minimizes a weighted sum of the root-mean-square fitting error and the $\ell_{1}$ norm of $\mathbf{\Phi}$
	\begin{equation}
	\mathbf{\Phi}^{\text{CS}} = {\arg \min}_\mathbf{\Phi} \; \| \mathbf{\Phi} \|_1 + \frac{\mu}{2} \| \mathbf{F} -  \mathbb{A} \mathbf{ \Phi} \|^2_2, 
	\label{eq:L1min}
	\end{equation}
where $\mathbf{F}$ is a vector composed of atomic forces and $\mathbb{A}$ is a matrix formed by the products the atomic displacements. The parameter $\mu$ adjusts the relative weights of the fitting error versus the absolute magnitude of the nonzero IFC components represented by the term with the $\ell_{1}$ norm; small values of $\mu$ favor solutions with very few nonzero IFCs at the expense of the accuracy of the fitted forces (``underfitting''), while very large $\mu$ will give a dense solution with many large nonzero IFCs that fits the DFT forces well, but may have poor predictive accuracy due to numerical noise, both random and systematic (``overfitting''). There is an optimal range of $\mu$ values between these two extremes where a sparse IFC vector $\mathbf{\Phi}$ can be obtained with excellent predictive accuracy. We note that the addition of the $\ell_1$ term solves both difficulties of the least-squares fitting approach described above. In practice, the optimal $\mu$ is determined by monitoring the predictive relative error of $\mathbf{F}$ for a leave-out subset of the training data not used in fitting, the details of which can be found in Ref.~\onlinecite{csld,Nelson2013}. In this study, since we are concerned with lattice heat transport properties at room temperature and only consider three-phonon processes, we extracted IFCs up to 3$^{\text{rd}}$-order using random small atomic displacements ranging from 0.01 to 0.04 \r{A} without considering higher-order IFCs. Fig.~\ref{fig:fitcsld} (a) shows the predictive relative force error for a leave-out subset as a function of $\mu$ for PbTe, which displays a rapid decease with $\mu$ increased from 10$^{-4}$ to 10$^{2}$ and very slight increase with $\mu$ further increased to 10$^{4}$. As a result, the optimal $\mu$ has a relatively large range from 10$^2$ to 10$^3$, in which we found computed lattice thermal conductivity to vary within 1.0\%. Fig.~\ref{fig:fitcsld} (b) shows the comparison between predicted forces and those directly computed by DFT (not used in fitting), further verifying the accuracy of fitted IFCs.

\subsection{Phonon dispersions and lattice thermal conductivity of PbTe using experimental lattice constant}
\begin{figure}[htp]
	\includegraphics[width = 0.5\linewidth]{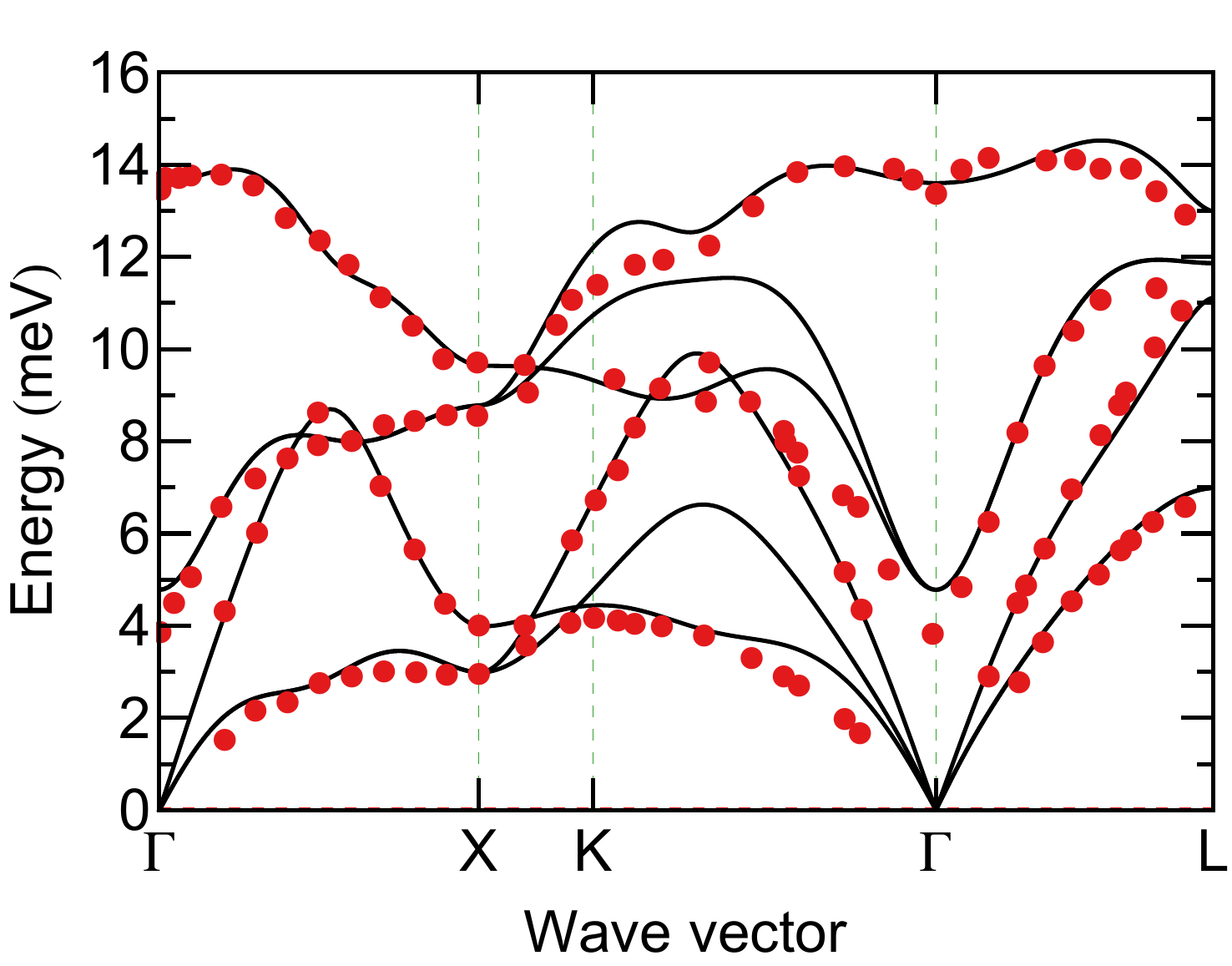}
	\caption{
	Computed phonon dispersions (black solid lines) using experimental lattice constant at 300 K compared with experimental data\cite{Cochran1966} (solid red disks) at the same temperature. 
	}
	\label{fig:CompDispPbTe}
\end{figure}

\begin{figure}[htp]
	\includegraphics[width = 0.5\linewidth]{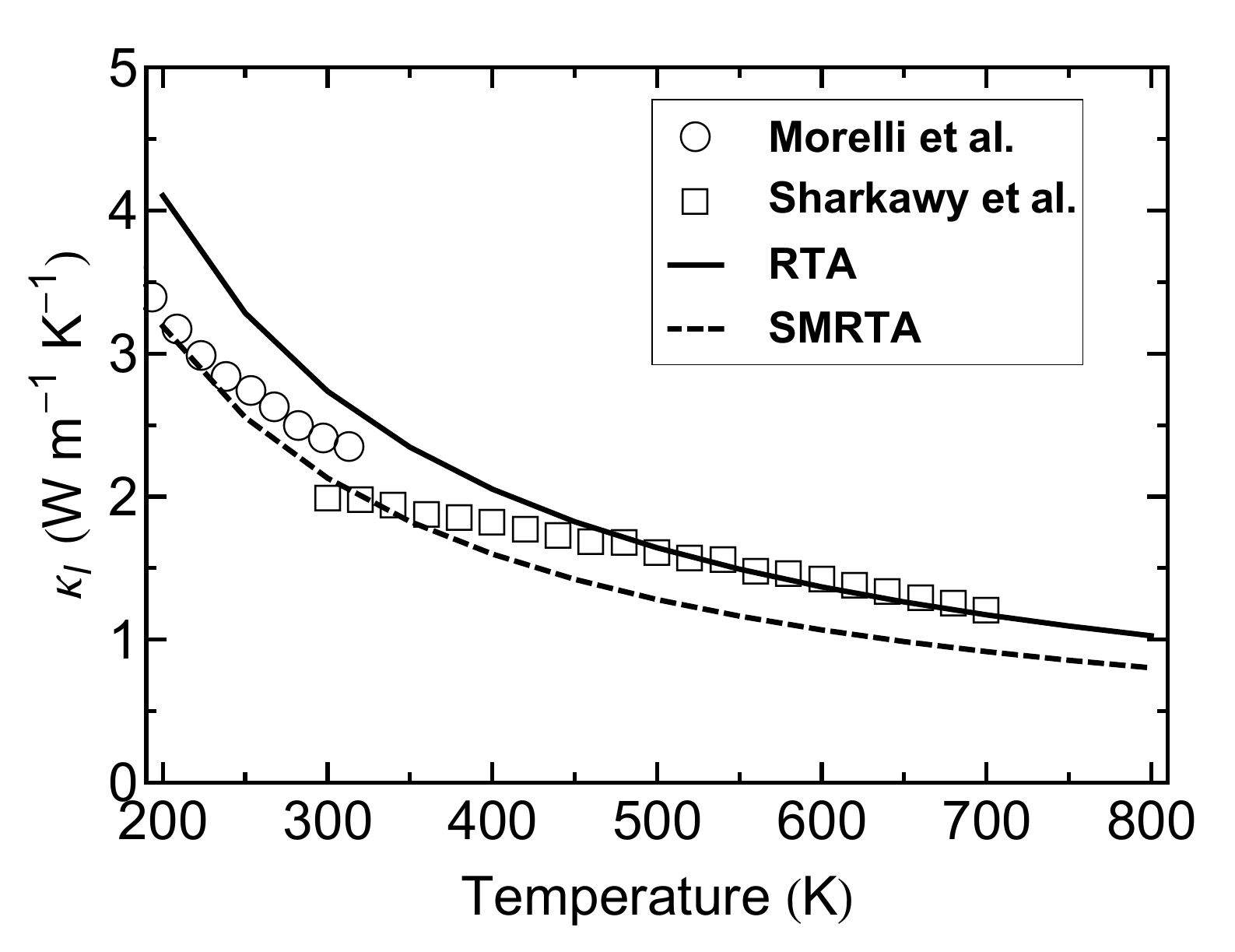}
	\caption{
	Computed lattice thermal conductivity using relaxation time approximation (RTA: solid lines) and single mode relaxation time approximation (SMRTA: dashed lines) compared with experimental measurements on single crystals from \textcite{Morelli2008} and polycrystalline from \textcite{Sharkawy1983}.
	}
	\label{fig:CompKappaPbTe}
\end{figure}

To further verify our results for PbTe, we benchmarked the computed phonon dispersions and lattice thermal conductivity of PbTe with available experimental measurements, as shown in Fig.~\ref{fig:CompDispPbTe} and Fig.~\ref{fig:CompKappaPbTe}. We found that the computed phonon dispersions agree well with experimental data for both acoustic branches and high-energy optical modes, except the zone-center low-lying transverse optical (TO) mode, where a slight overestimation is observed. Since it has been widely discussed and accepted that the softening of TO mode would lead to reduced lattice thermal conductivity through increasing scattering phase space,\cite{Skelton2014,Lee:2014aa} we expect that our overestimation of energy of TO mode leads to overestimated lattice thermal conductivity. Indeed, as shown in Fig.~\ref{fig:CompKappaPbTe}, our computed lattice thermal conductivity of PbTe at 300 K is larger than experimental values. 

The deviation from $1/T$ behavior in the thermal conductivity of PbTe may be attributed to (1) thermal expansion, (2) temperature-induced phonon renormalization (i.e. changes in phonon frequencies and phonon-phonon interaction strengths due to the increased ionic displacements at high temperature), and (3) significant contributions from beyond-three-phonon processes. Thermal expansion and phonon renormalization has been investigated in Refs.~\onlinecite{Skelton2014,Romero2015}. The focus of the current work, however, is to identify the alloying effects in PbTe. An investigations of the combined effects of alloying, strain, beyond-third-order phonon interactions, thermal expansion, and phonon renormalization is beyond the scope of the current study. Therefore, we limit our discussion mainly to 300 K for both pristine PbTe and PbTe alloyed with MgTe, CaTe, SrTe and BaTe. We expect that our main conclusions will remain qualitatively correct at higher temperatures. The detailed comparison with previous calculations and experiments are summarized in Table\ \ref{table:pPbTe} and \ref{table:MPbTe}.

\begin{table}[ht]
\caption{Lattice thermal conductivity (in W/m$\cdot$K) of pristine PbTe at 300 K, as computed by single mode relaxation time approximation (SMRTA), iterative solution of the Boltzmann Transport Equation (BTE), and as measured experimentally.} 
\centering 
\begin{tabular}{c | c | c | c } 
\hline\hline 
 & Computed (SMRTA) & Computed (Iterative BTE) & Experimental \\ [0.5ex] 
\hline 
Pristine PbTe  & 2.1 (this work) &  2.7 (this work) & 2.4\cite{Morelli2008} (single crystal) \\
              & 1.9\cite{zhiting} &                  & 2.0\cite{Sharkawy1983} (polycrystalline) \\
              & 2.1\cite{Skelton2014} &                   &  \\ [1ex] 
\hline 
\end{tabular}
\label{table:pPbTe} 
\end{table}

\begin{table}[ht]
\caption{Lattice thermal conductivity (in W/m$\cdot$K) of Pb$_{0.94}$M$_{0.06}$Te (M=Mg, Ca, Sr and Ba) at 300 K, as measured experimentally and as computed without and with strain induced by M substitution.} 
\centering 
\begin{tabular}{c | c | c | c | c} 
\hline\hline 
 & Pb$_{0.94}$Mg$_{0.06}$Te  & Pb$_{0.94}$Ca$_{0.06}$Te & Pb$_{0.94}$Sr$_{0.06}$Te & Pb$_{0.94}$Ba$_{0.06}$Te \\ [0.5ex] 
\hline 
Experimental & 1.74\cite{Zhao2013} & 1.33\cite{Biswas2011EES} & 1.98\cite{Tan2016}  & 1.2\cite{Biswas2011EES} (Pb$_{0.97}$Ba$_{0.03}$Te)  \\ [1ex] 
Computed (w/o strain) & 1.39 & 1.49 & 1.84 & 2.30 \\ [1ex] 
Computed (w/ strain) & 1.55  & 1.50 & 1.72 & 1.67 \\ [1ex] 
\hline 
\end{tabular}
\label{table:MPbTe} 
\end{table}

\clearpage
\newpage

\end{document}